\documentclass[amssymb,aps,pre,showpacs,manuscript,secnumarabic,showkeys]{revtex4}
\usepackage{amsmath}
\usepackage{graphicx}
\usepackage{amssymb}
\usepackage{esint}
\usepackage{mathrsfs}

\begin{document}

\title{Noise-Intensity Fluctuation in Langevin Model and\\
 Its Higher-Order Fokker--Planck Equation}

\author{Yoshihiko Hasegawa}

\email{hasegawa@cb.k.u-tokyo.ac.jp}

\affiliation{Department of Biophysics and Biochemistry, Graduate School of Science,
The University of Tokyo, Tokyo 113-0033, Japan}

\author{Masanori Arita}

\affiliation{Department of Biophysics and Biochemistry, Graduate School of Science,
The University of Tokyo, Tokyo 113-0033, Japan}

\affiliation{Institute for Advanced Biosciences, Keio University, Yamagata 997-0035,
Japan}

\date{November 26, 2010}
\begin{abstract}
In this paper, we investigate a Langevin model subjected to \emph{stochastic
intensity noise} (SIN), which incorporates temporal fluctuations in
noise-intensity. We derive a higher-order Fokker--Planck equation
(HFPE) of the system, taking into account the effect of SIN by the
adiabatic elimination technique. Stationary distributions of the HFPE
are calculated by using the perturbation expansion. We investigate
the effect of SIN in three cases: (a) parabolic and quartic bistable
potentials with additive noise, (b) a quartic potential with multiplicative
noise, and (c) a stochastic gene expression model. We find that the
existence of noise intensity fluctuations induces an intriguing phenomenon
of a bimodal-to-trimodal transition in probability distributions.
These results are validated with Monte Carlo simulations. 
\end{abstract}



\maketitle

\section{Introduction\label{sec:introduction}}

\begin{figure}
\begin{centering}
\includegraphics[width=6cm]{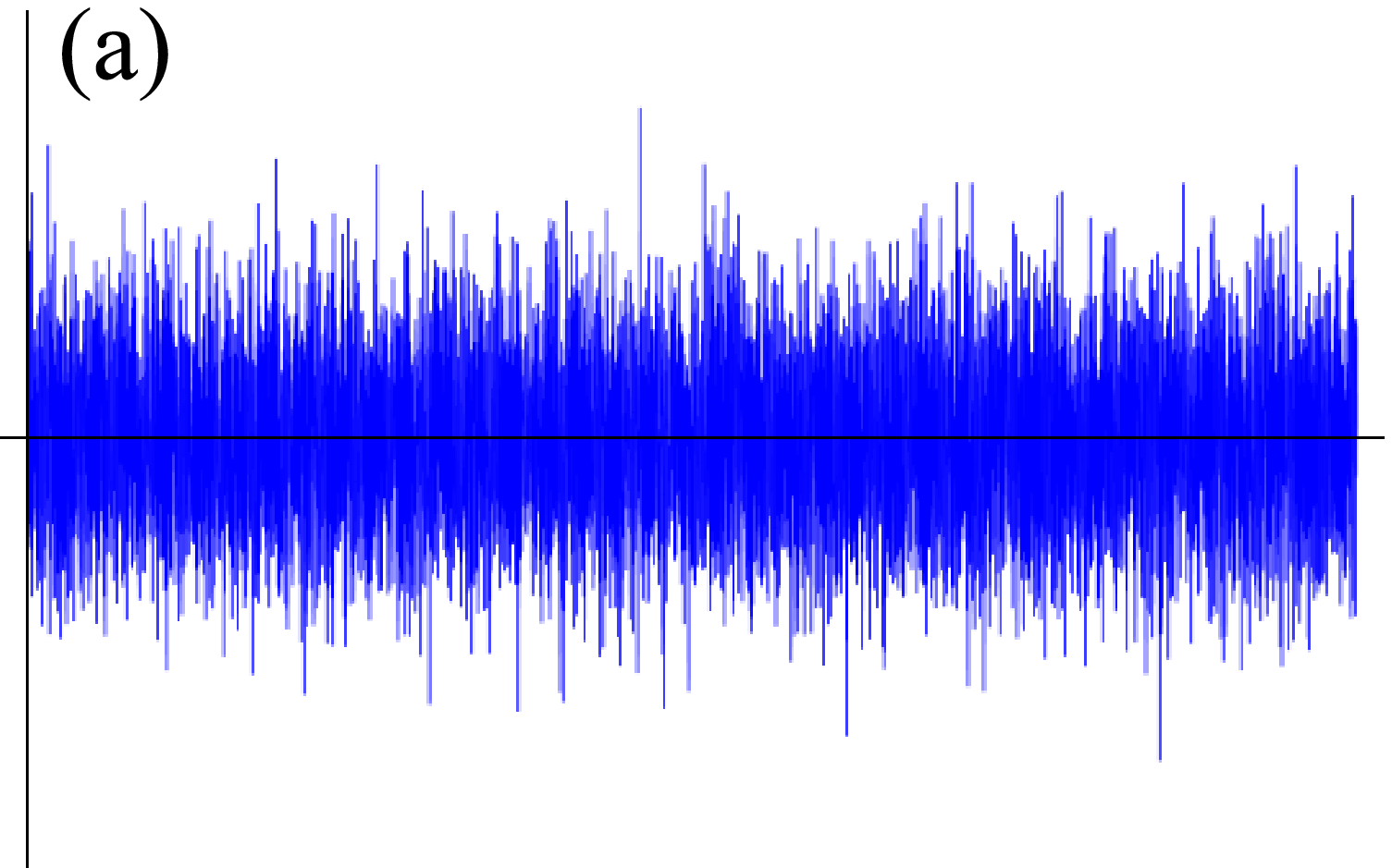}~~~\includegraphics[width=6cm]{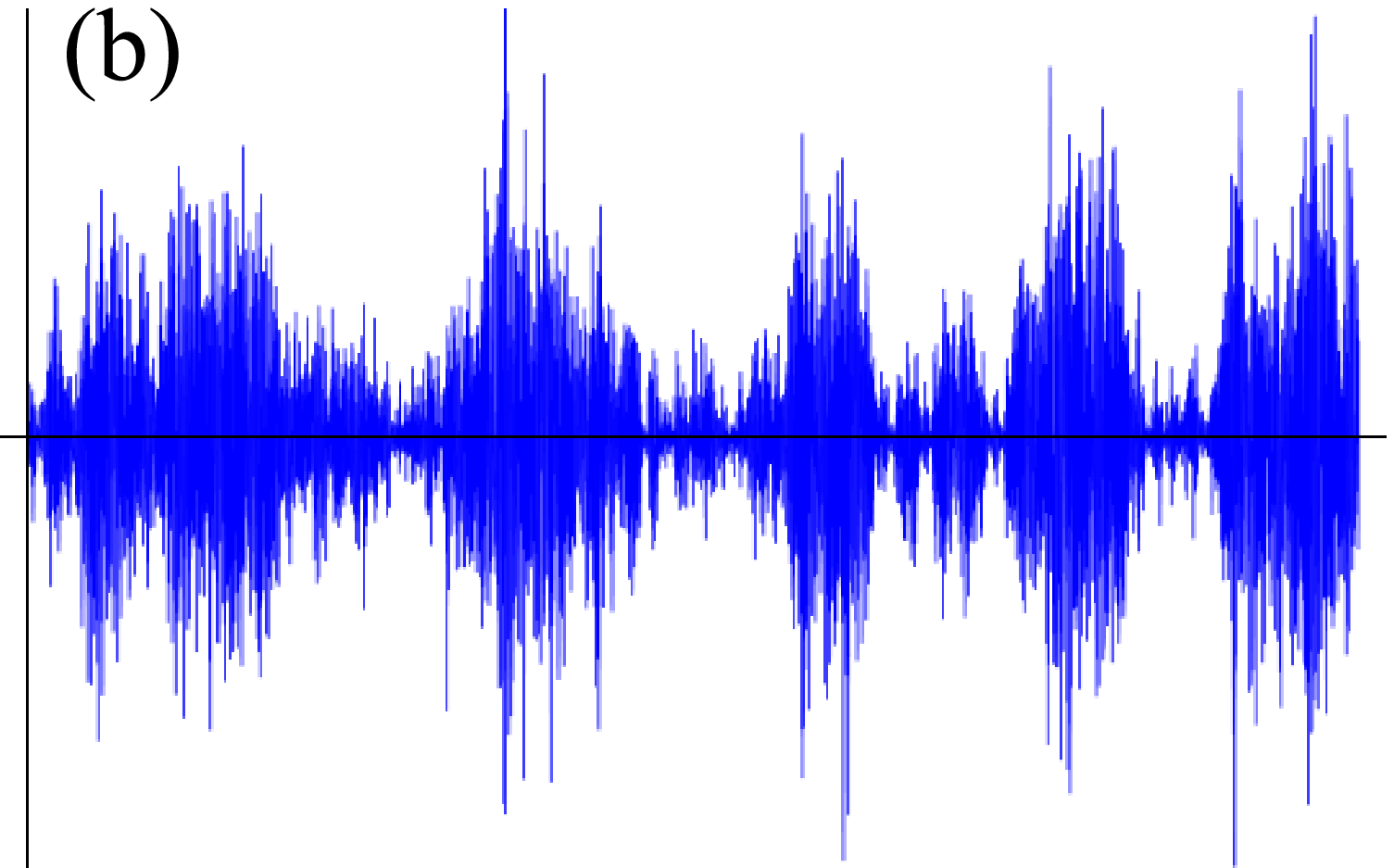} 
\par\end{centering}

\caption{(Color online) Intuitive paths of (a) white noise and (b) SIN whose
intensity is governed by the Ornstein--Uhlenbeck process {[}Eq.~(\ref{eq:shifted_OU}){]}.
\label{fig:intuitive_path}}

\end{figure}

Many real-world systems are inhomogeneous and fluctuant. Stochastic
processes are often used for modeling such fluctuant systems in many
fields, including physics, biology, and chemistry. The dynamics in
these systems can be described by a Langevin equation given by \begin{equation}
\frac{dx}{dt}=f(x)+g(x)\xi_{x}(t),\label{eq:basic_Langevin}\end{equation}
 where $f(x)=-\partial_{x}U(x)$, $U(x)$ denotes a potential, $g(x)$
is an arbitrary function of $x$, and $\xi_{x}(t)$ is the white Gaussian
noise with correlation $\left\langle \xi_{x}(t)\xi_{x}(t^{\prime})\right\rangle =2D_{x}\delta(t-t')$.
Although white noise can reflect microscale properties of fluctuations,
it is uniform when seen from mesoscopic or macroscopic time scales
{[}Fig.~\ref{fig:intuitive_path}(a){]}. One widely used approach
for incorporating mesoscopic or macroscopic inhomogeneity in noise
is colored noise, where the white noise $\xi_{x}(t)$ in Eq.~(\ref{eq:basic_Langevin})
is replaced by $z(t)$ with the Ornstein--Uhlenbeck process: \begin{equation}
\frac{dz}{dt}=-\frac{z}{\tau}+\frac{\xi_{z}(t)}{\tau},\label{eq:colored_OU}\end{equation}
 where $\xi_{z}(t)$ expresses the white Gaussian noise {[}$\left\langle \xi_{z}(t)\xi_{z}(t^{\prime})\right\rangle =2D_{z}\delta(t-t')${]}.
Equation~(\ref{eq:colored_OU}) yields colored noise with the finite
correlation time $\tau$, that is, $\langle z(t)z(t^{\prime})\rangle=(D_{z}/\tau)\exp\left\{ -\vert t-t^{\prime}\vert/\tau\right\} $.
The existence of correlation time in the noise sources can induce
many phenomena, such as resonant activation \cite{Doering:1992:ResonantActivation}
and noise-enhanced stability \cite{Mantegna:1996:NES,Dubkov:2004:NES,Spagnolo:2004:NES,Spagnolo:2008:NES_review,Fiasconaro:2009:NESinColoredNoise}.

In the present paper, we deal with mesoscopic time-scale inhomogeneity
in a way other than with colored noise; here, we consider temporal
noise-intensity fluctuations. We assume that the noise intensity in
Eq.~(\ref{eq:basic_Langevin}) is not constant but modulated by the
Ornstein--Uhlenbeck process. Our model is described by the following
coupled Langevin equations instead of Eq.~(\ref{eq:basic_Langevin}):
\begin{eqnarray}
\frac{dx}{dt} & = & f(x)+g(x)s\xi_{x}(t),\label{eq:basic_Langevin_SIN}\\
\frac{ds}{dt} & = & -\gamma(s-\alpha)+\sqrt{\gamma}\xi_{s}(t).\label{eq:shifted_OU}\end{eqnarray}
 Here, $\xi_{s}(t)$ is the white Gaussian noise {[}$\left\langle \xi_{s}(t)\xi_{s}(t^{\prime})\right\rangle =2D_{s}\delta(t-t')${]},
and $\gamma$ and $\alpha$ are the relaxation rate and the mean of
the Ornstein--Uhlenbeck process, respectively. The intensity-modulated
noise term $s\xi_{x}(t)$ in Eq.~(\ref{eq:basic_Langevin_SIN}) is
herein called \emph{stochastic intensity noise} (SIN) {[}Fig. 1(b){]}.
This point of view was originally introduced in Heston's stochastic
volatility models in financial engineering \cite{Heston:1993:Volatility}
and has since been analyzed in econophysics \cite{Dragulescu:2002:Heston}.
With $f(x)\propto x$ and $g(x)\propto x$, Eqs.~(\ref{eq:basic_Langevin_SIN})
and~(\ref{eq:shifted_OU}) are similar to those in the Heston model,
where the variance of noise is governed by the Feller process (also
referred to as the square-root process or the Cox--Ingersoll--Ross
process). Escape events in the Heston model were studied in Ref.~\cite{Bonanno:2007:EscapeSV}
using a cubic potential. Note that the variable $s$ in Eqs.~(\ref{eq:basic_Langevin_SIN})
and~(\ref{eq:shifted_OU}) takes negative as well as positive values
since our model can be considered as white noise with multiplicative
term $s$, which is in contrast with the positive variance in the
Heston model \cite{Heston:1993:Volatility}. In physics, superstatistics
includes the concept of temporal and/or spatial environmental fluctuations
\cite{Wilk:2000:NEXTParam,Beck:2001:DynamicalNEXT,Beck:2003:Superstatistics,Beck:2006:SS_Brownian,Beck:2009:RecentSS}.
Superstatistics was originally introduced, under specific conditions,
to account for asymptotic power-law distributions (e.g., $q$-exponential
distributions and $q$-Gaussian distributions) that emerge as maximizers
of non-additive (Tsallis) entropy \cite{Tsallis:1988:Generalization,Tsallis:2009:NonextensiveBook}.
Superstatistics has since been applied to the interpretation of quasi-equilibrium
thermodynamics, and concepts of superstatistics have also been applied
to stochastic processes \cite{Beck:2001:DynamicalNEXT,Beck:2006:SS_Brownian,Jizba:2008:SuposPD,Queiros:2008:SSMultiplicative,Hasegawa:2010:qExpBistable,Rodriguez:2007:SS_Brownian}.
In particular, Ref.~\cite{Jizba:2008:SuposPD} extended superstatistics
to the path-integral representation and showed that some stochastic
models can be covered by this representation. A direct connection
between Tsallis statistics and financial stochastic processes was
indicated in Ref.~\cite{Queiros:2005:VolatilityNEXT}.

In many biological and chemical processes, the relaxation time of
$x$ may be larger than that of environmental fluctuations (noise-intensity
processes). This is the case in stochastic gene expression models
in which the decay time of $x$ is on the order of minutes \cite{Liu:2004:FluctuationGeneTrans}
(Sec.~\ref{sub:gene_expression}). Bearing this fact in mind, we
will investigate systems driven by SIN with faster decay time compared
with that of $x$ ($\gamma\gg1$). Furthermore, in real-world systems,
we expect that the $f(x)$ in Eq.~(\ref{eq:basic_Langevin_SIN})
is often given by a complex nonlinear function and is also accompanied
by nontrivial multiplicative noise expressed by an appropriate $g(x)$.
These properties are different from those for financial engineering.
In order to obtain the probability distribution $P(x,t)$, we use
adiabatic elimination with eigenfunction expansion \cite{Kaneko:1976:AdiabaticElim}.
We obtain a higher-order Fokker--Planck equation (HFPE) with higher-order
derivatives, not included in the conventional Fokker--Planck equation
(FPE). We calculate stationary distributions of the systems described
by the HFPE by using the perturbation expansion.

To investigate the effects of SIN, we consider stationary distributions
for three cases: (a) parabolic and quartic bistable potentials with
additive noise, (b) a quartic potential with multiplicative noise
and (c) a stochastic gene expression model. In the additive noise
case (a), we show that SIN changes the stationary distribution from
exponential forms (the Boltzmann--Gibbs distribution) to non-exponential
forms. At the same time, the stationary distributions of case (a)
are sharpened because of noise-intensity fluctuations. In case (b),
we show that the existence of noise-intensity fluctuations induces
the transition of distributions: the stationary distributions are
uni-, bi-, or trimodal, depending on the SIN parameters. It is important
to note that the trimodal distribution does not emerge under white
noise. In case (c) which we call the gene expression model, a nonlinear
function is given as a drift term (change in expression levels). Thus,
case (c) shows that our approximation scheme can be applied to general
configurations that include non-trivial drift terms. 

The remainder of this paper is organized as follows: In Sec.~\ref{sec:model},
we describe the model proposed in this paper. Adiabatic elimination
with eigenfunction expansion is applied to the model in Sec.~\ref{sec:adiabatic_elimination}
(details of the derivation of the HFPE are explained in the Appendix).
In Sec.~\ref{sec:stationary_dist}, we proceed to the calculation
of the stationary distributions of the obtained HFPE by using of the
perturbation expansion. In Sec. \ref{sec:experiments}, we investigate
effects of SIN in the three cases (a), (b), and (c) mentioned above.
In Sec.~\ref{sec:discussion}, we analyze effects of higher-order
derivatives in the HFPE on positivity and moments of distribution
functions, and also consider the opposite case, in which a decay time
of the noise-intensity fluctuations is very slow ($\gamma\rightarrow0$).
Finally, we give the conclusions in Sec.~\ref{sec:conclusion}.

\section{The Model\label{sec:model}}

We consider the Langevin equations given by Eqs.~(\ref{eq:basic_Langevin_SIN})
and~(\ref{eq:shifted_OU}). Since Eq.~(\ref{eq:shifted_OU}) does
not depend on $x$, the stationary distribution $P_{st}(s)$ of $s$
is easily obtained: \begin{equation}
P_{st}(s)=\frac{1}{\sqrt{2\pi D_{s}}}\exp\left\{ -\frac{(s-\alpha)^{2}}{2D_{s}}\right\} .\label{eq:y_stationary}\end{equation}
 Time evolution of the probability distribution $P(x,s;t)$ is given
by\begin{equation}
\frac{\partial}{\partial t}P(x,s;t)=\mathrm{L}_{\mathrm{FP}}(x,s)P(x,s;t),\label{eq:FPE}\end{equation}
 where $\mathrm{L}_{\mathrm{FP}}(x,s)$ is a FPE operator composed
of \begin{equation}
\mathrm{L}_{\mathrm{FP}}(x,s)=\mathrm{L}_{x}(x,s)+\gamma\mathrm{L}_{s}(s),\label{eq:FPE_operator}\end{equation}
 with \begin{equation}
\mathrm{L}_{x}(x,s)=-\frac{\partial}{\partial x}\left\{ f(x)+D_{x}g(x)g^{\prime}(x)s^{2}\right\} +D_{x}s^{2}\frac{\partial^{2}}{\partial x^{2}}g(x)^{2},\label{eq:Lxs}\end{equation}
 \begin{equation}
\mathrm{L}_{s}(s)=\frac{\partial}{\partial s}(s-\alpha)+D_{s}\frac{\partial^{2}}{\partial s^{2}}.\label{eq:Ls}\end{equation}
 We employ Stratonovich's calculus because it is expected to be more
relevant to physical applications than It\^o's \cite{Horsthemke:1984:NITBook}.
We calculate a projected time evolution equation of $P(x;t)=\int ds\, P(x,s;t)$
that can be applied to general configurations.

\section{Elimination of the Fast Variable\label{sec:adiabatic_elimination}}

In this section, we eliminate $s$ from Eq.~(\ref{eq:FPE}) by using
adiabatic elimination under the assumption that $\gamma\gg1$ (\emph{i.e.},
the decay time of $s$ is much faster than that of $x$). Adiabatic
elimination has its origin in the Born--Oppenheimer approximation.
Because of its usefulness, adiabatic elimination has been applied
to many stochastic systems \cite{Gardiner:1984:AEinStocproc1,Jung:1987:ColoreApproximation}.
In stochastic volatility models, $\gamma\rightarrow0$ adiabatic elimination
has been applied \cite{Remer:2004:HestonAE,Biro:2008:nonGaussAE}.
These models take advantage of the fact that noise intensity (volatility)
changes on a macroscopic time scale in financial markets (this description
agrees with Beck's superstatistical Brownian motion \cite{Beck:2006:SS_Brownian}).
In these studies, the obtained solutions do not explicitly include
$\gamma$ as a parameter. In the present paper, we consider the $\gamma\gg1$
case and derive a HFPE containing $O(\gamma^{-1})$ terms by using
adiabatic elimination with eigenfunction expansion \cite{Kaneko:1976:AdiabaticElim}.

By applying adiabatic elimination up to $O(\gamma^{-1})$, we obtain
the following HFPE (see the Appendix for the derivation): \begin{equation}
\frac{\partial}{\partial t}P(x;t)=\left\{ -\frac{\partial}{\partial x}f(x)+Q\triangle_{g}+\frac{R}{\gamma}\triangle_{g}^{2}\right\} P(x;t),\label{eq:KM_expansion_ori}\end{equation}
 with \begin{equation}
\triangle_{g}=\frac{\partial^{2}}{\partial x^{2}}g(x)^{2}-\frac{\partial}{\partial x}g^{\prime}(x)g(x),\label{eq:def_op}\end{equation}
 where $Q$ and $R$ in Eq.~(\ref{eq:KM_expansion_ori}) are defined
as follows:\[
Q=D_{x}\left(D_{s}+\alpha^{2}\right),\]
 \[
R=D_{x}^{2}D_{s}\left(4\alpha^{2}+D_{s}\right).\]
 For the case of additive noise {[}\emph{i.e.}, $g(x)=1${]}, Eq.~(\ref{eq:KM_expansion_ori})
reduces to \begin{equation}
\frac{\partial}{\partial t}P(x;t)=\left\{ -\frac{\partial}{\partial x}f(x)+Q\frac{\partial^{2}}{\partial x^{2}}+\frac{R}{\gamma}\frac{\partial^{4}}{\partial x^{4}}\right\} P(x;t).\label{eq:KM_expansion}\end{equation}
 Because Eqs.~(\ref{eq:KM_expansion_ori}) and (\ref{eq:KM_expansion})
have derivatives of orders higher than two, Eqs.~(\ref{eq:KM_expansion_ori})
and (\ref{eq:KM_expansion}) are referred to as the HFPE. In systems
driven by colored noise, adiabatic elimination up to $O(\tau)$ can
be expressed by the conventional FPE \cite{Kaneko:1976:AdiabaticElim}
{[}see Eq.~(\ref{eq:colored_OU}){]}. On the other hand, $O(\gamma^{-1})$
terms are accompanied by non-FPE terms in our SIN case. Therefore,
in order to incorporate the effect of $\gamma$, we must use the HFPE
form of Eq.~(\ref{eq:KM_expansion_ori}). For $\gamma^{-1}\rightarrow0$,
the last term of Eq.~(\ref{eq:KM_expansion_ori}) vanishes, and we
obtain FPE: \begin{equation}
\frac{\partial}{\partial t}P(x;t)=-\frac{\partial}{\partial x}\left\{ f(x)+Qg^{\prime}(x)g(x)\right\} P(x;t)+Q\frac{\partial^{2}}{\partial x^{2}}g(x)^{2}P(x;t).\label{eq:trivial_FPE}\end{equation}
 Eq.~(\ref{eq:trivial_FPE}) corresponds to conventional ($\gamma\rightarrow\infty$)
adiabatic elimination. $Q$ plays the role of the effective noise
intensity of the corresponding white Gaussian noise process.

\section{Stationary Distribution\label{sec:stationary_dist}}

In many practical cases, stationary distributions play important roles.
In this section, we calculate the stationary distribution $P_{st}(x)$
of Eq.~(\ref{eq:KM_expansion_ori}), which yields the following differential
equation:\begin{equation}
-\frac{f}{Q}P_{st}(x)+\left(\frac{\partial}{\partial x}g^{2}-g^{\prime}g\right)P_{st}(x)+\varepsilon\left(\frac{\partial}{\partial x}g^{2}-g^{\prime}g\right)\frac{\partial}{\partial x}\left(\frac{\partial}{\partial x}g^{2}-g^{\prime}g\right)P_{st}(x)=0,\label{eq:Pst_DE}\end{equation}
 where \begin{equation}
\varepsilon=\frac{R}{\gamma Q}=\frac{D_{x}D_{s}(4\alpha^{2}+D_{s})}{\gamma(D_{s}+\alpha^{2})}.\label{eq:eps_dependence}\end{equation}
 It is easy to see that a solution of Eq.~(\ref{eq:Pst_DE}) for
$\varepsilon=0$ is the stationary distribution of Eq.~(\ref{eq:trivial_FPE}).
Thus, for $\varepsilon\ll1$, it is expected that a solution of Eq.~(\ref{eq:Pst_DE})
can be approximated by the perturbation from the stationary distribution
for $\varepsilon=0$. In solving Eq.~(\ref{eq:Pst_DE}), we adopt
the perturbation expansion \cite{Kubo:1973:FlucMacrovariable,Horthemke:1980:PerturbationExpansion,Knessl:1984:WKB,Frank:2005:DelayFPE},
given by\[
P_{st}(x)=\Pi_{0}(x)+\varepsilon\Pi_{1}(x)+\varepsilon^{2}\Pi_{2}(x)+\cdots.\]
 We specifically define the following truncated first-order approximation:
\begin{equation}
P_{st}^{(1)}(x)=\Pi_{0}(x)+\varepsilon\Pi_{1}(x),\label{eq:WKB_expansion}\end{equation}
 where $\Pi_{0}(x)$ is the stationary distribution of the unperturbed
case {[}Eq.~(\ref{eq:trivial_FPE}){]}, and $\Pi_{1}(x)$ corresponds
to the first-order correction term. Substituting Eq.~(\ref{eq:WKB_expansion})
into Eq.~(\ref{eq:Pst_DE}) and comparing the order of $\varepsilon$
up to $O(\varepsilon)$, we obtain \begin{eqnarray}
O(1) &  & \left\{ \frac{f(x)}{Q}+g^{\prime}(x)g(x)\right\} \Pi_{0}(x)=\frac{\partial}{\partial x}g(x)^{2}\Pi_{0}(x),\label{eq:WKB_DE1}\\
O(\varepsilon) &  & \left\{ \frac{f(x)}{Q}+g^{\prime}(x)g(x)\right\} \Pi_{1}(x)=\frac{\partial}{\partial x}g(x)^{2}\Pi_{1}(x)+\phi(x),\label{eq:WKB_DE2}\end{eqnarray}
 where\[
\phi(x)=\left\{ \frac{\partial}{\partial x}g(x)^{2}-g^{\prime}(x)g(x)\right\} \frac{\partial}{\partial x}\left\{ \frac{\partial}{\partial x}g(x)^{2}-g^{\prime}(x)g(x)\right\} \Pi_{0}(x).\]
 From Eq.~(\ref{eq:WKB_DE1}), $\Pi_{0}(x)$ is given by\[
\Pi_{0}(x)=\frac{1}{Z|g(x)|}\exp\left(\frac{1}{Q}\int^{x}dv\,\frac{f(v)}{g(v)^{2}}\right),\]
 where $Z$ is a normalizing term {[}$\int dx\,\Pi_{0}(x)=1${]}.
Since Eq.~(\ref{eq:WKB_DE2}) is a first-order differential equation,
$\Pi_{1}(x)$ can be obtained analytically in many cases, as will
be discussed in Sec.~\ref{sec:experiments}.

The perturbation expansion yields reliable results for $\varepsilon\ll1$.
Combining the approximation condition of adiabatic elimination and
the perturbation expansion, it can be concluded that we are able to
calculate the stationary distribution by Eq.~(\ref{eq:WKB_expansion})
for systems with $\gamma\gg1$ and $\varepsilon\ll1$.

\section{Model Calculations\label{sec:experiments}}

We apply the approximation method obtained in Sec.~\ref{sec:stationary_dist}
to three cases: parabolic and quartic bistable potentials with additive
noise (Sec.~\ref{sub:p_q_additive}), a quartic potential with multiplicative
noise (Sec.~\ref{sub:quartic_mult}), and a stochastic gene expression
model (Sec.~\ref{sub:gene_expression}). We analyze effects of noise-intensity
fluctuations on shapes of the stationary distributions. We also carried
out Monte Carlo (MC) simulations to determine the reliability of our
approximation. In the following, we show results of three calculation
methods: 
\begin{itemize}
\item \textbf{HFPE stationary distribution}\\
 The stationary distribution of the HFPE {[}Eq.~(\ref{eq:KM_expansion_ori}){]}
is shown using the perturbation expansion explained in previous sections
including the correction term $\Pi_{1}(x)$. This is an $O(\gamma^{-1})$
approximation. 
\item \textbf{FPE stationary distribution}\\
 The stationary distribution of the FPE {[}Eq.~(\ref{eq:trivial_FPE}){]}
is shown. This is identical to $\Pi_{0}(x)$ and does not include
the effect of noise-intensity fluctuations. This case corresponds
to $\gamma^{-1}\rightarrow0$ adiabatic elimination. 
\item \textbf{MC simulation}\\
 We have employed a simple Euler-forward scheme with a time step
size $\Delta t=10^{-6}$ (for details of the algorithm, readers may
refer to Sec. 3.6 of Ref.~\cite{Risken:1989:FPEBook}). We have used
$N=5\times10^{7}$ points to calculate empirical distributions. For
all the line-symmetric potentials, MC data are symmetrized with respect
to $x=0$. 
\end{itemize}
For evaluating quantitatively the reliability of HFPE stationary distributions,
we calculate the root-mean-square (RMS) distance between the HFPE
(or FPE) and MC distributions. We first divide an interval of $x$
(window of each figure) into $M$ points ($M=100$), each of which
is denoted by $x_{i}$. Let $\widehat{P}_{i}$ and $P_{st}(x_{i})$
be density values of the MC and HFPE (FPE) distributions at $x_{i}$,
respectively. RMS is defined by \[
\mathrm{RMS}=\sqrt{\frac{1}{M}\sum_{i=1}^{M}\left(\widehat{P}_{i}-P_{st}(x_{i})\right)^{2}}.\]

\subsection{Parabolic and Quartic Potentials with Additive Noise\label{sub:p_q_additive}}

We investigate effects of additive SIN in parabolic and quartic bistable
potentials. We first calculate the stationary distribution of the
parabolic potential {[}$U(x)=x^{2}/2${]} with additive SIN {[}$g(x)=1${]}.
According to Eqs.~(\ref{eq:WKB_DE1}) and~(\ref{eq:WKB_DE2}), $\Pi_{0}(x)$
and $\Pi_{1}(x)$ are given as follows: \begin{equation}
\Pi_{0}(x)=\sqrt{\frac{1}{2\pi Q}}\exp\left(-\frac{x^{2}}{2Q}\right),\label{eq:quadra_Pi0}\end{equation}
 \begin{equation}
\Pi_{1}(x)=\sqrt{\frac{1}{2\pi Q}}\left(\frac{x^{4}}{4Q^{3}}-\frac{3x^{2}}{2Q^{2}}\right)\exp\left(-\frac{x^{2}}{2Q}\right)+C\exp\left(-\frac{x^{2}}{2Q}\right),\label{eq:quadra_Pi1}\end{equation}
 where $C$ is determined by the normalization condition {[}\emph{$\int dx\, P_{st}^{(1)}(x)=1$}{]}:
\[
C=\frac{3}{4\sqrt{2\pi}Q^{3/2}}.\]
 Substituting Eqs.~(\ref{eq:quadra_Pi0}) and~(\ref{eq:quadra_Pi1})
in Eq.~(\ref{eq:WKB_expansion}), we obtain the stationary distribution
$P_{st}^{(1)}(x)$.

\begin{figure}
\begin{centering}
\includegraphics[width=8cm]{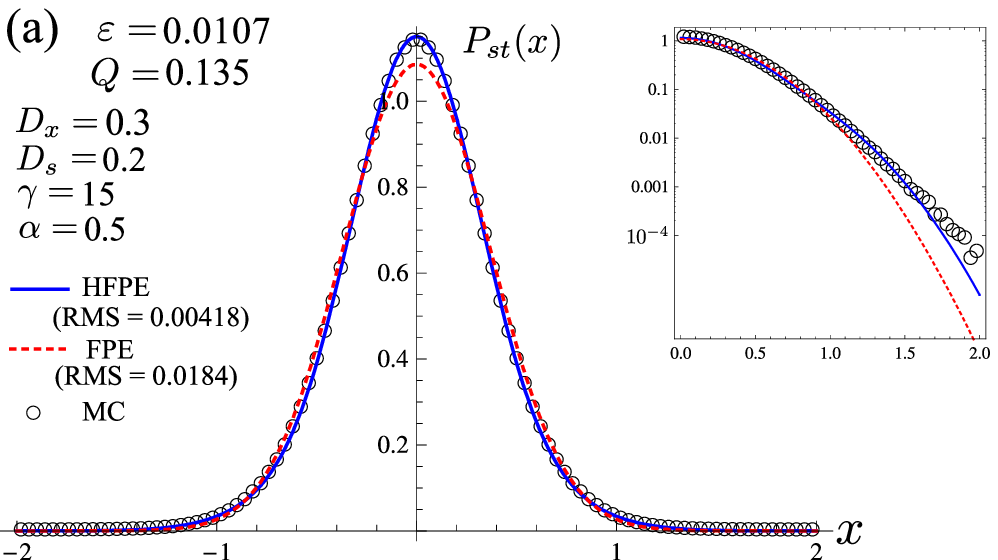} 
\par\end{centering}

\begin{centering}
\includegraphics[width=7cm]{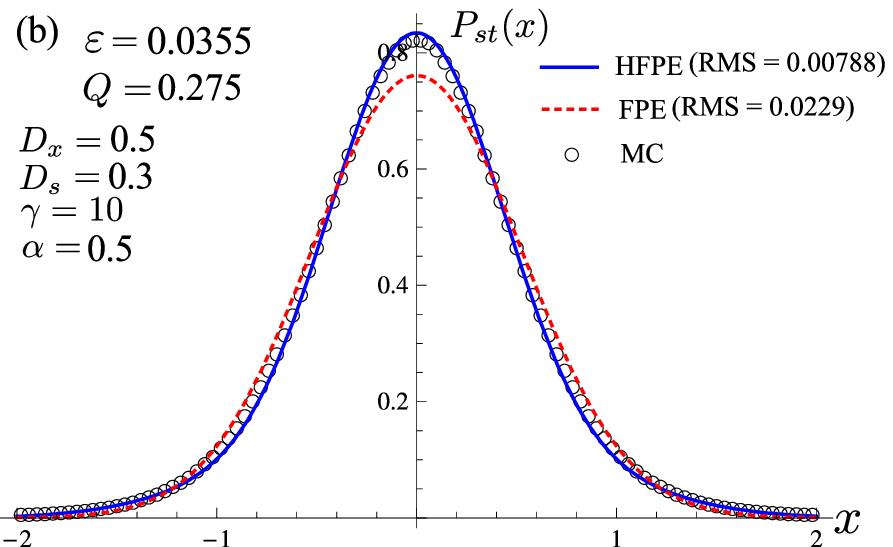}~~~~\includegraphics[width=7cm]{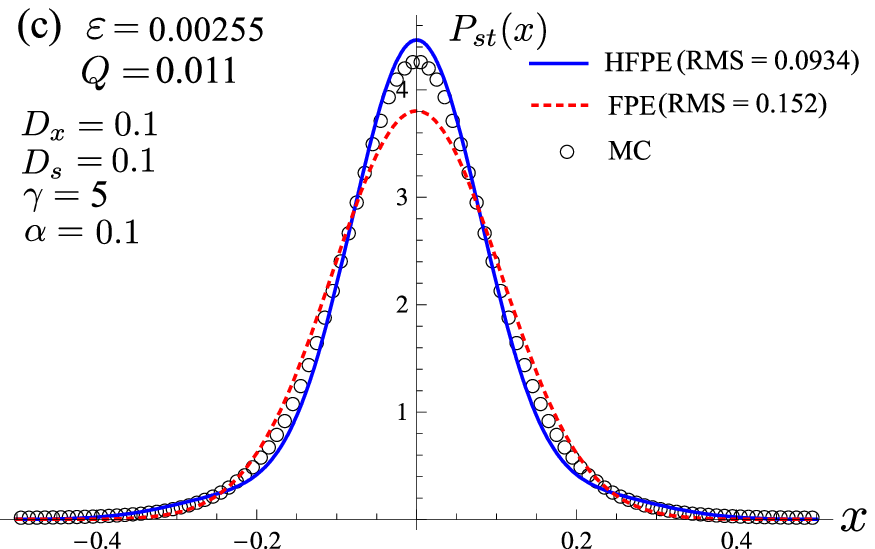} 
\par\end{centering}

\caption{(Color online) Stationary distributions for the parabolic potential
with additive noise calculated by HFPE (solid lines), FPE (dashed
lines), and MC (circles). Parameter values are (a) $D_{x}=0.3$, $D_{s}=0.2$,
$\gamma=15$, and $\alpha=0.5$ ($Q=0.135$, $\varepsilon=0.0107$);
(b) $D_{x}=0.5$, $D_{s}=0.3$, $\gamma=10$, and $\alpha=0.5$ ($Q=0.275$,
$\varepsilon=0.0355$); and (c) $D_{x}=0.1$, $D_{s}=0.1$, $\gamma=5$,
and $\alpha=0.1$ ($Q=0.011$, $\varepsilon=0.00255$). The inset
in (a) is plotted on a log scale ($x\ge0$) and the others on linear
scales. \label{fig:quadratic_comp}}

\end{figure}

Figure~\ref{fig:quadratic_comp} shows stationary distributions for
the parabolic potential case calculated by HFPE (solid lines), FPE
(dashed lines), and MC simulation (circles). We also show the RMS
distance between the HFPE (FPE) and MC distributions in Fig.~\ref{fig:quadratic_comp}(a),
(b), and (c). We see that the densities of the HFPE at $x=0$ are
higher than those of FPE under the existence of noise-intensity fluctuations.
Because the stationary distributions with noise-intensity fluctuations
are approximately realized as superposition of many Gaussian distributions
with different variances, those with smaller variance make the stationary
distributions sharper. It is important to note that the stationary
distributions of the HFPE (solid lines) are not Gaussian. This non-Gaussianity
is derived from the existence of higher-order derivatives than the
second in the HFPE (see Sec.~\ref{sub:higher_terms}). From the RMS
values, we see that the distance between the HFPE and MC distributions
is smaller than that between FPE and MC, indicating a better agreement
of HFPE distributions. This result supports the reliability of our
approximation scheme. The HFPE stationary distribution of $\gamma=15$
{[}Fig.~\ref{fig:quadratic_comp}(a){]} exhibits better agreement
than the $\gamma=5$ case {[}Fig.~\ref{fig:quadratic_comp}(c){]}
because larger $\gamma$ yields better approximation for the adiabatic
elimination technique. The inset in Fig.~\ref{fig:quadratic_comp}(a)
shows a log-scale plot for the $x\ge0$ region, showing that the HFPE
stationary distribution has fatter tails, which in turn implies that
the existence of noise-intensity fluctuations makes distributions
fatter (see Sec.~\ref{sub:higher_terms}). Furthermore, the HFPE
stationary distribution exhibits better agreement with the MC simulations
than the FPE in tail areas, indicating that our approximation scheme
offers reliable results even in tail areas for sufficiently large
$\gamma$ and sufficiently small $\varepsilon$.

We next calculate the stationary distributions of the quartic bistable
potential {[}$U(x)=x^{4}/4-x^{2}/2${]} driven by additive SIN {[}$g(x)=1${]}.
Because bistable potentials can represent switching dynamics, they
are very important in many fields. $\Pi_{0}(x)$ and $\Pi_{1}(x)$
are given by\[
\Pi_{0}(x)=\frac{1}{Z}\exp\left\{ -\frac{1}{Q}\left(\frac{x^{4}}{4}-\frac{x^{2}}{2}\right)\right\} ,\]
 \begin{eqnarray*}
\Pi_{1}(x) & = & \frac{x^{2}}{ZQ^{3}}\left\{ 3Q^{2}-\frac{3}{2}\left(x^{2}-1\right)^{2}Q+x^{2}\left(\frac{x^{6}}{10}-\frac{3x^{4}}{8}+\frac{x^{2}}{2}-\frac{1}{4}\right)\right\} \\
 &  & \times\exp\left\{ -\frac{1}{Q}\left(\frac{x^{4}}{4}-\frac{x^{2}}{2}\right)\right\} +C\exp\left\{ -\frac{1}{Q}\left(\frac{x^{4}}{4}-\frac{x^{2}}{2}\right)\right\} ,\end{eqnarray*}
 with \[
Z=\frac{\pi}{2}\exp\left(\frac{1}{8Q}\right)\left\{ I_{-\frac{1}{4}}\left(\frac{1}{8Q}\right)+I_{\frac{1}{4}}\left(\frac{1}{8Q}\right)\right\} ,\]
 where $I_{n}(z)$ is the modified Bessel function of the first kind
and $C$ is to be numerically evaluated by the normalization condition.

\begin{figure}
\begin{centering}
\includegraphics[width=9cm]{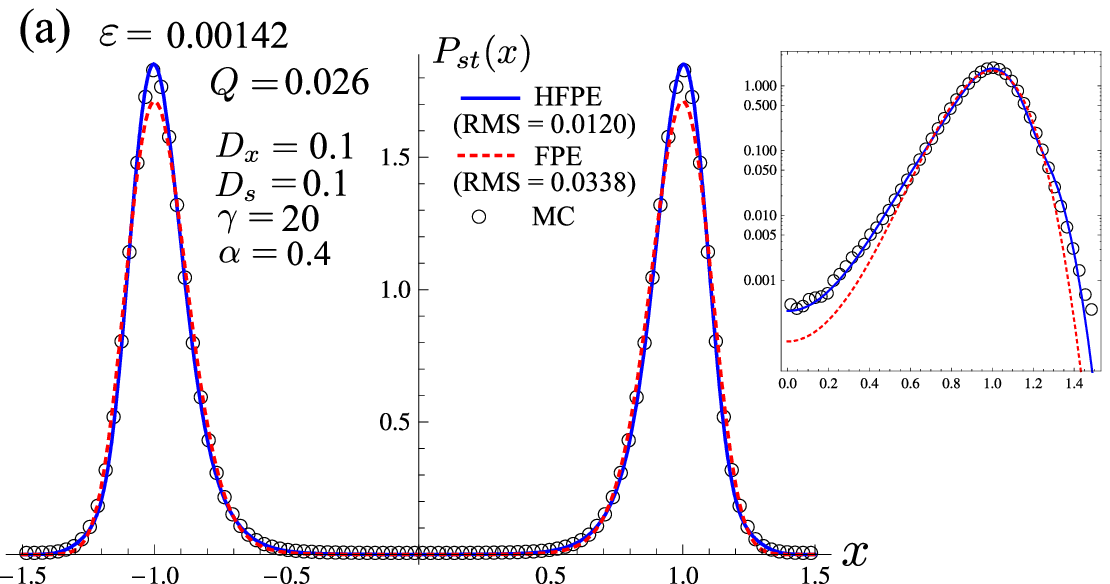} 
\par\end{centering}

\begin{centering}
\includegraphics[width=7cm]{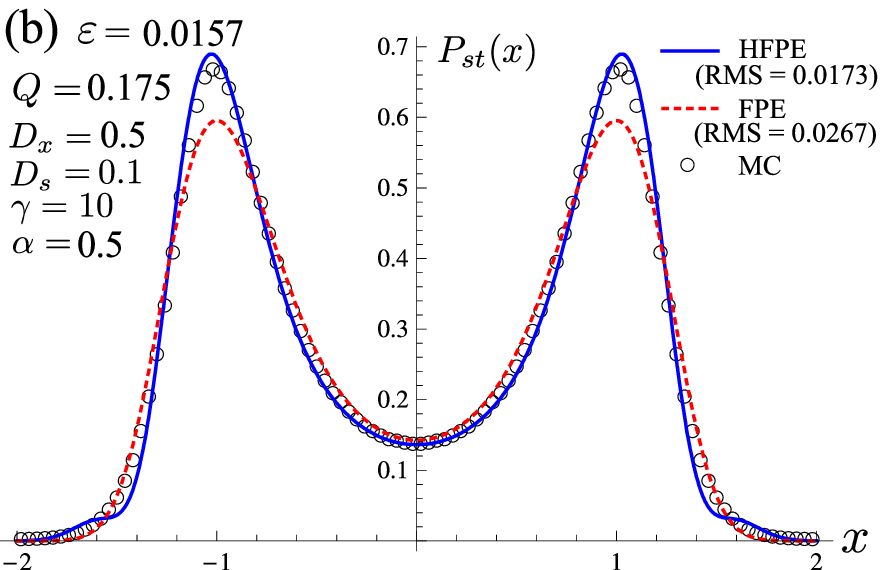}~~~~\includegraphics[width=7cm]{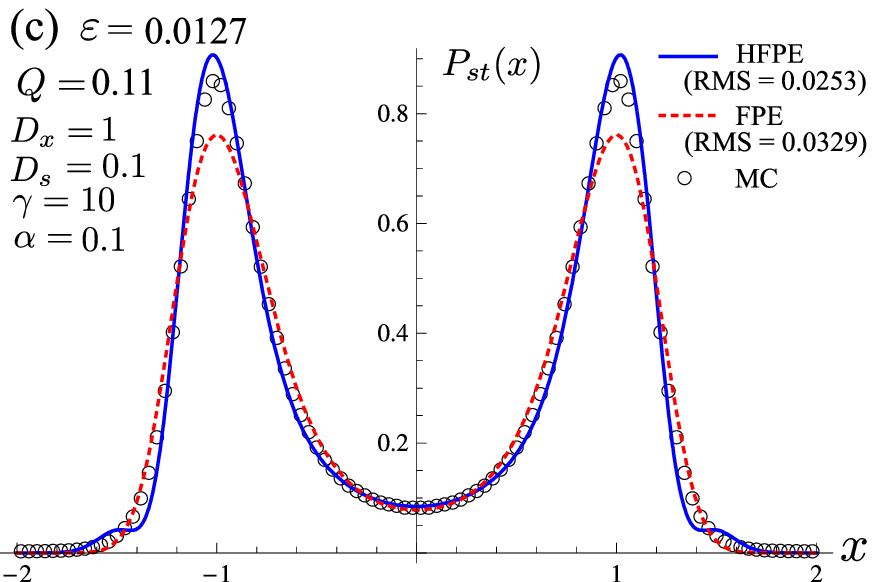} 
\par\end{centering}

\caption{(Color online) Stationary distributions of the quartic bistable potential
with additive noise calculated by HFPE (solid lines), FPE (dashed
lines), and MC (circles). Parameter values are (a) $D_{x}=0.1$, $D_{s}=0.1$,
$\gamma=20$, and $\alpha=0.4$ ($Q=0.026$, $\varepsilon=0.00142$);
(b) $D_{x}=0.5$, $D_{s}=0.1$, $\gamma=10$, and $\alpha=0.5$ ($Q=0.175$,
$\varepsilon=0.0157$); and (c) $D_{x}=1$, $D_{s}=0.1$, $\gamma=10$,
and $\alpha=0.1$ ($Q=0.11$, $\varepsilon=0.0127$). The inset in
(a) is plotted on a log scale ($x\ge0$) and the others on linear
scales. \label{fig:quartic_comp}}

\end{figure}

Figure~\ref{fig:quartic_comp} shows the stationary distributions
calculated by the three methods for the quartic bistable potential
case. The meaning of each symbol is the same as in Fig.~\ref{fig:quadratic_comp}.
We also see that the densities of the HFPE at stable sites ($x=-1$
and $1$) are higher than those of the FPE under the existence of
noise-intensity fluctuations. The inset in Fig.~\ref{fig:quartic_comp}(a)
(a log-scale plot) shows that the distribution has fatter tails under
noise-intensity fluctuations. As in the case of the parabolic potential,
the stationary distributions under additive SIN are not Boltzmann--Gibbs
distributions. According to the RMS values, we also see a good agreement
between the HFPE and MC results. From the log-scale plot, the HFPE
result shows better agreement than the FPE also in tail areas for
sufficiently large $\gamma$ and sufficiently small $\varepsilon$.

\subsection{Quartic Bistable Potential with Multiplicative Noise\label{sub:quartic_mult}}

We consider the quartic bistable potential as in the previous section
{[}$U(x)=x^{4}/4-x^{2}/2${]}, but the system in this section is driven
by linear multiplicative noise {[}$g(x)=x${]}. Multiplicative noise
plays an important role in many phenomena. $\Pi_{0}(x)$ and $\Pi_{1}(x)$
are given by \begin{equation}
\Pi_{0}(x)=\frac{1}{Z}|x|^{\frac{1-Q}{Q}}\exp\left(-\frac{x^{2}}{2Q}\right),\label{eq:quartic_mult_un}\end{equation}

\begin{eqnarray*}
\Pi_{1}(x) & = & C|x|^{\frac{1-Q}{Q}}\exp\left(-\frac{x^{2}}{2Q}\right)+\frac{1}{ZQ^{3}}|x|^{\frac{1-Q}{Q}}\exp\left(-\frac{x^{2}}{2Q}\right)\\
 &  & \times\biggl\{\frac{x^{6}}{6}-\frac{3}{4}(2Q+1)x^{4}+\frac{1}{2}\left(4Q^{2}+6Q+3\right)x^{2}-\log|x|\biggr\},\end{eqnarray*}
 where $Z$ and $C$ are \[
Z=(2Q)^{\frac{1}{2Q}}\Gamma\left(\frac{1}{2Q}\right),\]
 \[
C=\frac{1}{12Q^{3}(2Q)^{\frac{1}{2Q}}\Gamma\left(\frac{1}{2Q}\right)}\left\{ 6\log\left(2Q\right)+6\psi\left(\frac{1}{2Q}\right)-4Q(Q+3)-11\right\} .\]
 Here, $\psi(x)$ is the digamma function {[}$\psi(x)=\partial_{x}\log\Gamma(x)${]}.
We see that $\Pi_{0}(x)$ is composed of two Gamma distributions for
$x>0$ and $x<0$. Depending on the effective noise intensity $Q$,
$\Pi_{0}(x)$ is unimodal ($Q\ge1$) or bimodal ($0<Q<1$).

\begin{figure}
\begin{centering}
\includegraphics[width=8cm]{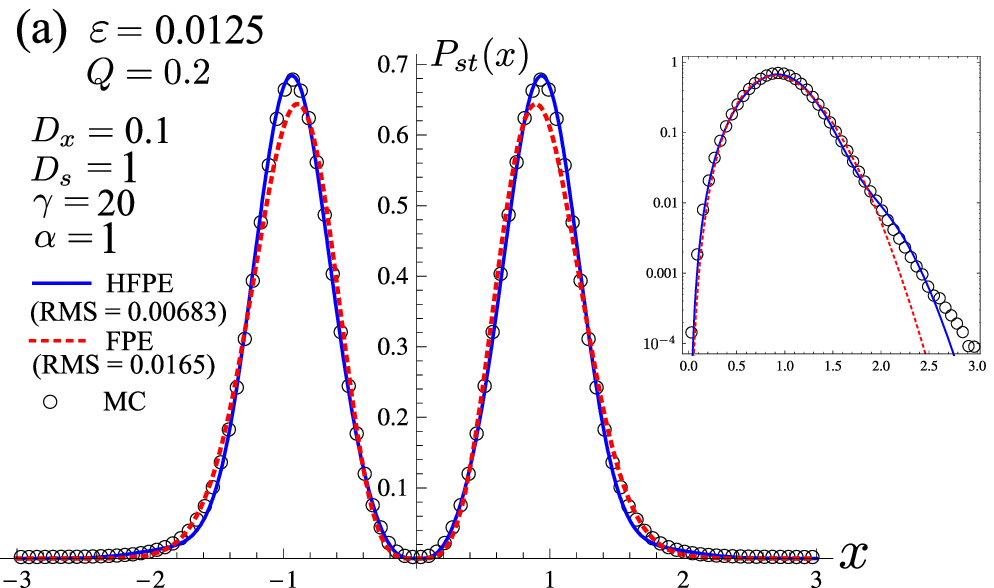} 
\par\end{centering}

\begin{centering}
\includegraphics[width=7cm]{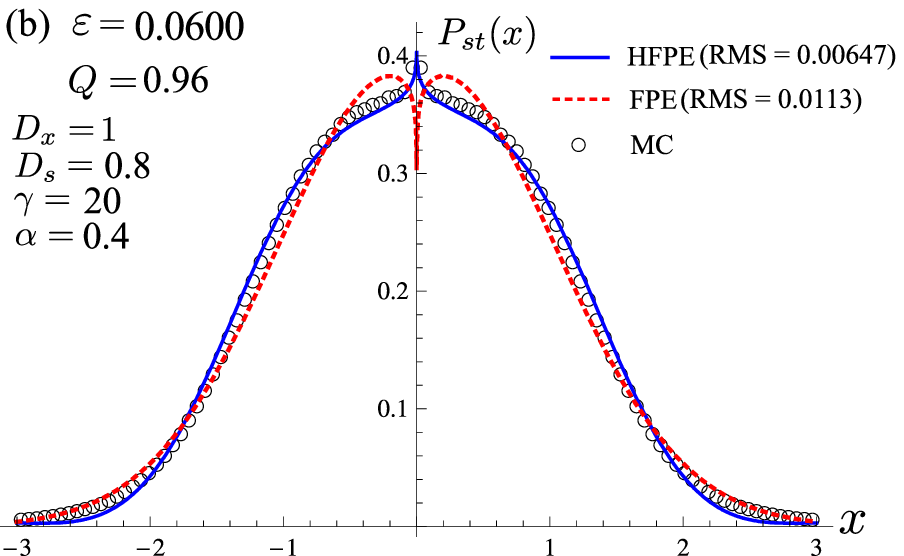}~~~~\includegraphics[width=7cm]{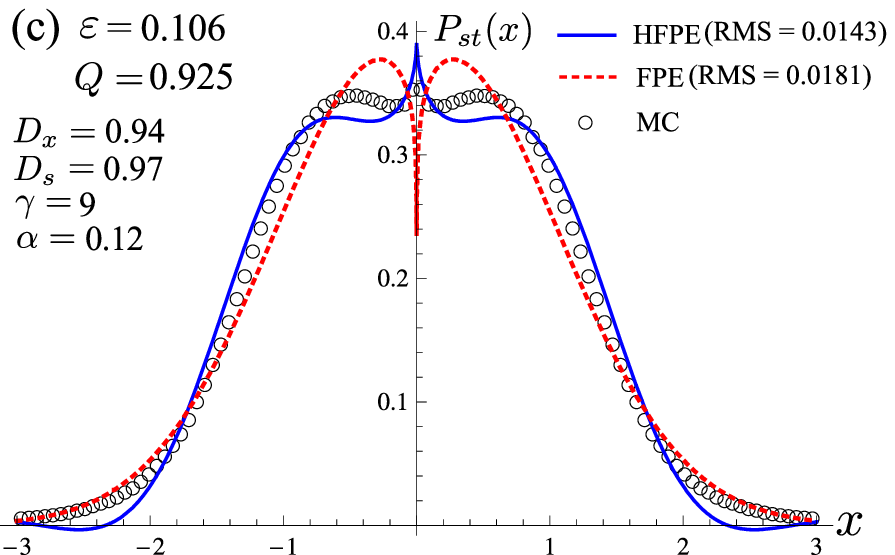} 
\par\end{centering}

\caption{(Color online) Stationary distributions of the quartic bistable potential
with linear multiplicative noise {[}$g(x)=x${]} calculated by HFPE
(solid lines), FPE (dashed lines), and MC (circles). Parameter values
are (a) $D_{x}=0.1$, $D_{s}=1$, $\gamma=20$, and $\alpha=1$ ($Q=0.2$,
$\varepsilon=0.0125$); (b) $D_{x}=1$, $D_{s}=0.8$, $\gamma=20$,
and $\alpha=0.4$ ($Q=0.96$, $\varepsilon=0.06$); and (c) $D_{x}=0.94$,
$D_{s}=0.97$, $\gamma=9$, and $\alpha=0.12$ ($Q=0.925$, $\varepsilon=0.106$).
The inset in (a) is plotted on a log scale ($x\ge0$) and the others
on linear scales. \label{fig:quartic_mult_comp}}

\end{figure}

Figure~\ref{fig:quartic_mult_comp} represents the probability densities
of three typical cases. Figure~\ref{fig:quartic_mult_comp}(a) shows
the bimodal stationary distribution ($Q<1$). In this case, the existence
of noise-intensity fluctuations makes densities around $x=1$ and
$x=-1$ higher. This is also observed in the additive noise case.
Figure~\ref{fig:quartic_mult_comp}(b) shows qualitatively different
shapes between stationary distributions of FPE and HFPE. Because $Q=0.96<1$
in Fig.~\ref{fig:quartic_mult_comp}(b), the FPE stationary distribution
(dashed line) exhibits bimodality. On the other hand, the stationary
distribution of HFPE is unimodal (this result agrees with the MC simulation).
This indicates that the transition from bimodal to unimodal is induced
by the existence of noise-intensity fluctuations, and the FPE stationary
distribution does not correctly reflect this property. Figure~\ref{fig:quartic_mult_comp}(c)
also shows the qualitatively different results between FPE and HFPE
stationary distributions. It is interesting to see that the HFPE stationary
distribution is trimodal, which is not observed without the existence
of noise-intensity fluctuations (the MC simulation also exhibits trimodality).
This trimodal distribution is a result of superposition of unimodal
and bimodal distributions. This result shows that the HFPE formulation
is essential to understand systems driven by multiplicative noise.
In all figures, the RMS values show that the HFPE distributions provide
more reliable results than FPE distributions. From the inset in Fig.~\ref{fig:quartic_mult_comp}(a)
(a log-scale plot), we see that the GPFE distribution agrees with
the MC distribution for tail areas for sufficiently large $\gamma$
and sufficiently small $\varepsilon$.

\subsection{Gene Expression Model\label{sub:gene_expression}}

\begin{figure}
\begin{centering}
\includegraphics[width=7cm]{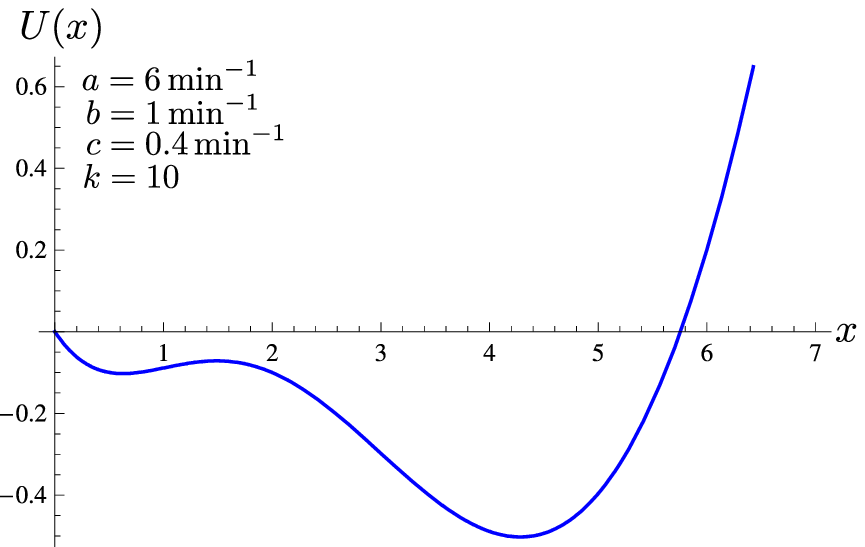} 
\par\end{centering}

\caption{(Color online) A potential function $U(x)$ for the gene expression
model with $a=6\min^{-1}$, $b=1\min^{-1}$, $c=0.4\min^{-1}$, and
$k=10$. $U(x)$ exhibits bistability with these parameters. \label{fig:gene_potential}}

\end{figure}

Langevin equations are often used for describing stochastic chemical
reactions. In order to show that our approximation scheme can be applied
to systems including non-trivial drift terms, we apply it to a stochastic
gene expression model \cite{Smolen:1998:GeneRegulation}. A simple
genetic expression model with self-regulation is given by \begin{equation}
\frac{dx}{dt}=\frac{ax^{2}}{x^{2}+k}-bx+c,\label{eq:gene_expression}\end{equation}
 where $x$ is the concentration of a transcription factor (protein),
and $a$, $b$, $c$, and $k$ are model parameters taking positive
values (for details, see Ref.~\cite{Smolen:1998:GeneRegulation}).
In this model, the potential function is given by \begin{equation}
U(x)=a\sqrt{k}\arctan\left(\frac{x}{\sqrt{k}}\right)+\frac{bx^{2}}{2}-(a+c)x.\label{eq:gene_potential}\end{equation}
 The units of $a$, $b$, and $c$ are $\min^{-1}$, implying that
the decay time of Eq.~(\ref{eq:gene_expression}) is on the order
of minutes. Because gene expression goes through various fluctuations,
some of the fluctuations have faster decay times than $x$. Then the
model can be cast in the form of Langevin equation (\ref{eq:basic_Langevin_SIN})
with \begin{equation}
f(x)=\frac{ax^{2}}{x^{2}+k}-bx+c,\label{eq:gene_drift}\end{equation}
 \[
g(x)=1.\]
 Specifically, we employed $a=6\min^{-1}$, $b=1\min^{-1}$, $c=0.4\min^{-1}$,
and $k=10$, as in Ref.~\cite{Liu:2004:FluctuationGeneTrans}. With
the adopted parameter values, Eq.~(\ref{eq:gene_potential}) exhibits
bistability (Fig.~\ref{fig:gene_potential}). It is considered that
this bistability is responsible for the genetic switch.

Using our approximation method, $\Pi_{0}(x)$ and $\Pi_{1}(x)$ are
given as follows: \begin{equation}
\Pi_{0}(x)=\frac{1}{Z}\exp\left\{ -\frac{U(x)}{Q}\right\} ,\label{eq:gene_Pi0}\end{equation}

\begin{eqnarray}
\Pi_{1}(x) & = & C\exp\left(-\frac{U(x)}{Q}\right)-\frac{1}{8Q^{3}Z}\exp\left(-\frac{U(x)}{Q}\right)\nonumber \\
 &  & \times\biggl[-2b^{3}x^{4}+8b^{2}(a+c)x^{3}+12b\left(-a^{2}-2ca-c^{2}+bQ\right)x^{2}\nonumber \\
 &  & +8\left\{ a^{3}+3ca^{2}+3\left(c^{2}-b(bk+Q)\right)a+c^{3}-3bcQ\right\} x\nonumber \\
 &  & +\frac{3ak}{x^{2}+k}\left\{ 3xa^{2}+4(bk-2Q+cx)a-8Q(c-bx)\right\} \nonumber \\
 &  & -3a\sqrt{k}\left(5a^{2}+12ca+8c^{2}-8b^{2}k\right)\arctan\left(\frac{x}{\sqrt{k}}\right)\nonumber \\
 &  & +24ab(a+c)k\log\left(x^{2}+k\right)+\frac{2ak}{\left(x^{2}+k\right)^{2}}\left(-kxa^{2}+6kQa+8Q^{2}x\right)\biggr],\label{eq:gene_Pi1}\end{eqnarray}
 where $Z$ and $C$ are normalizing constants, which are numerically
evaluated so that $\int dx\,\Pi_{0}(x)=1$ and $\int dx\, P_{st}^{(1)}(x)=1$.
Although the concentration $x$ cannot take negative values, the obtained
stationary distribution has a very small magnitude in the $x<0$ areas,
which can be ignored in practical calculations.

\begin{figure}
\begin{centering}
\includegraphics[width=7cm]{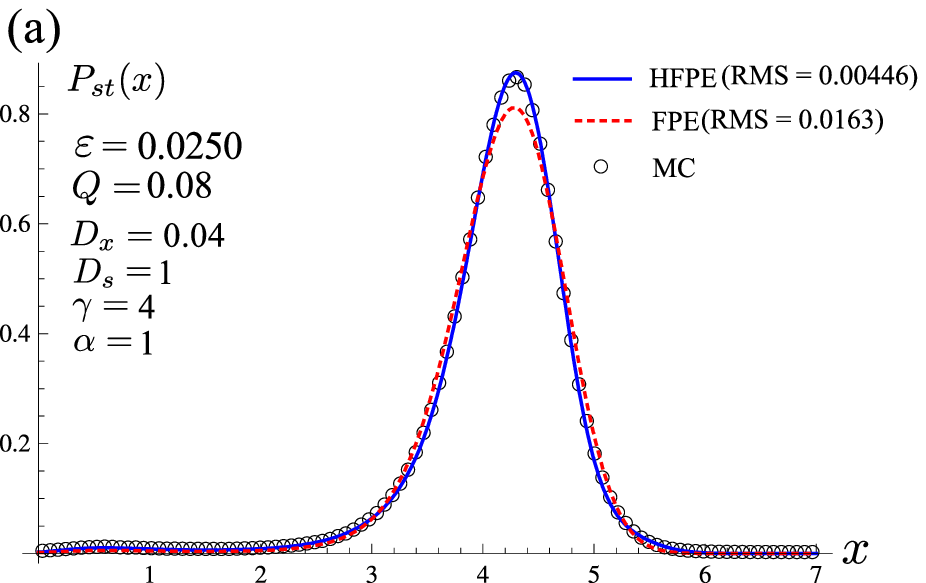}~~~~\includegraphics[width=7cm]{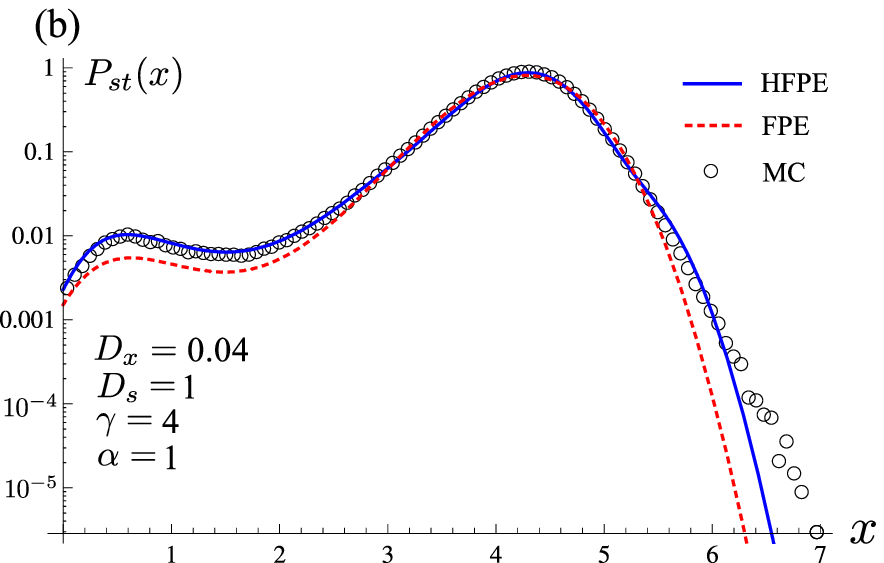} 
\par\end{centering}

\caption{(Color online) Stationary distributions of the stochastic gene expression
model calculated by HFPE (solid lines), FPE (dashed lines), and MC
(circles). Parameter values are $D_{x}=0.04$, $D_{s}=1$, $\gamma=4$,
and $\alpha=1$ ($Q=0.08$, $\varepsilon=0.025$). (a) is plotted
on a linear scale and (b) is on a log scale. \label{fig:gene_expression}}

\end{figure}

Figure~\ref{fig:gene_expression} represents probability distributions
calculated by the three methods. The distributions in (a) and (b)
are plotted on linear and log scales, respectively. We see that the
density of the HFPE (solid line) at the stable site ($x\simeq4.3$)
is higher than the case without noise-intensity fluctuations (dashed
line). From the RMS values in Fig.~\ref{fig:gene_expression}(a),
the HFPE stationary distribution agrees well with the MC simulations.
In the tail areas {[}Fig.~\ref{fig:gene_expression}(b){]}, the HFPE
result exhibits good agreement with the MC simulations, whereas the
FPE result does not. This result indicates that our approximation
method can be applied to systems with non-trivial drift terms.

\section{Discussion\label{sec:discussion}}

\subsection{Effects of higher-order derivative terms\label{sub:higher_terms}}

In the projected time evolution equations of our model, recursive
expansion leads to equations having an infinite number of terms. In
this paper, we have truncated at $O(\gamma^{-1})$, including derivatives
only up to fourth order. Pawula's theorem \cite{Pawula:1967:ApproxFPE}
indicates that to guarantee the positivity of the distribution functions,
the truncation must be after the first- or second-order term (this
corresponds to FPE). Otherwise, the equations must include an infinite
number of terms. However, it has been reported that time evolution
equations truncated at $n\ge3$ are practically meaningful \cite{Titulaer:1978:HighFricBrownian,Risken:1979:TruncatedFPE,Risken:1987:TruncatedKME}.
Indeed our approximation including derivatives of higher order than
two have shown better results than FPE solutions. In particular, it
has been shown in Figs.~\ref{fig:quartic_mult_comp}(b) and (c) that
stationary distributions of FPE and HFPE have qualitatively different
shapes, and our results using HFPE are supported by MC simulations.

Ref.~\cite{Plyukhin:2005:GFPE} gives an intuitive explanation of
the effect of higher-order derivatives, which are not included in
the conventional FPE, on the positivity of solutions. Following Ref.~\cite{Plyukhin:2005:GFPE},
we explain the parabolic potential case. Using Eq.~(\ref{eq:trivial_FPE}),
the FPE of the parabolic potential case is given by \begin{equation}
\frac{\partial}{\partial t}P(x;t)=\left\{ \frac{\partial}{\partial x}x+Q\frac{\partial^{2}}{\partial x^{2}}\right\} P(x;t).\label{eq:para_FPE}\end{equation}
 Let $x_{m}$ be points where $P(x;t)$ is locally minimal with respect
to $x$. According to requirements of the minima and the positivity
of $P(x_{m};t)$, $\partial_{t}P(x_{m};t)$ is always positive: \begin{equation}
\frac{\partial}{\partial t}P(x_{m};t)=x\frac{\partial}{\partial x}P(x;t)\biggr|_{x=x_{m}}+P(x_{m};t)+Q\frac{\partial^{2}}{\partial x^{2}}P(x;t)\biggr|_{x=x_{m}}>0.\label{eq:positive_timederiv}\end{equation}
 Eq.~(\ref{eq:positive_timederiv}) guarantees the positivity of
initially positive solutions because at the minima $P(x_{m};t)$ is
increasing as a function of $t$. On the other hand, the positivity
of $\partial_{t}P(x_{m};t)$ is not generally satisfied in HFPE cases,
since the fourth-order derivative term $(R/\gamma)\partial_{x}^{4}P(x;t)|_{x=x_{m}}$
can take any values. However, the effect of the fourth-order term
on the sign of $\partial_{t}P(x_{m};t)$ is negligible under sufficiently
smooth initial distributions and sufficiently large $\gamma$, since
the fourth-order term is on the order of $\gamma^{-1}$. From the
above explanation, it is generally expected that solutions of HFPE
can satisfy positivity under sufficiently large $\gamma$.

In order to analyze effects of higher-order derivatives on moments
for the additive noise case, we consider the following moment expansion
using Eq.~(\ref{eq:KM_expansion}): \begin{equation}
\frac{\partial}{\partial t}\left\langle x^{n}\right\rangle =n\left\langle f(x)x^{n-1}\right\rangle +Qn(n-1)\left\langle x^{n-2}\right\rangle +\frac{R}{\gamma}n(n-1)(n-2)(n-3)\left\langle x^{n-4}\right\rangle ,\label{eq:moment_expansion}\end{equation}
 where $\left\langle \cdots\right\rangle $ represents the expectation
with respect to $P(x;t)$ {[}$\left\langle A(x)\right\rangle =\int dx\, A(x)P(x;t)${]}.
We consider the stationary case of the parabolic potential case {[}$f(x)=-x${]}.
Equation~(\ref{eq:moment_expansion}) in this case is described by
\[
\left\langle x\right\rangle =0,\quad\left\langle x^{2}\right\rangle =Q,\quad\left\langle x^{3}\right\rangle =0,\quad\left\langle x^{4}\right\rangle =3Q^{2}+\frac{6R}{\gamma}.\]
 Kurtosis is given by\begin{equation}
\kappa=\frac{\left\langle x^{4}\right\rangle }{\left\langle x^{2}\right\rangle ^{2}}-3=\frac{6D_{s}\left(4\alpha^{2}+D_{s}\right)}{\gamma\left(D_{s}+\alpha^{2}\right)^{2}}.\label{eq:parabolic_kurtosis}\end{equation}
 $\gamma\rightarrow\infty$ yields the Gaussian stationary distribution
with $\kappa=0$. Equation~(\ref{eq:parabolic_kurtosis}) shows that
smaller $\gamma$ yields a stationary distribution with larger kurtosis.
This means that the distributions have fatter tails under noise-intensity
fluctuations. This result agrees with those of Sec.~\ref{sub:p_q_additive}.
In Ref.~\cite{Beck:2006:SS_Brownian}, fat-tailed stationary distributions
are derived not from time evolution equations but from a calculation
of the expectation with respect to a noise-intensity distribution.
Our results show that derivatives of orders higher than two are required
to yield fat-tailed stationary distributions in the parabolic potential.
Furthermore, the importance of the fourth-order derivative was pointed
out in Ref.~\cite{Rodriguez:2007:SS_Brownian}, which considered
superstatistical Brownian particles using mesoscopic nonequilibrium
thermodynamics \cite{Mazur:1999:MNET}. Ref.~\cite{Rodriguez:2007:SS_Brownian}
obtained a time evolution equation in the Fourier space. Inclusion
of a quartic term in the terms of a Fourier conjugate variable indicates
the existence of a fourth-order derivative in real space. In specific
cases, stationary distributions of their model are very close to a
$q$-Gaussian distribution, which are strongly connected to Beck's
superstatistics.

The stationary distributions of the FPE (Sec.~\ref{sub:p_q_additive})
belong to the exponential family, which are strongly connected to
the Boltzmann--Gibbs--Shannon (BGS) entropy. On the other hand, the
stationary distributions of the HFPE with fatter tails are not exponential.
Non-exponential distributions in statistical mechanics have been extensively
discussed in Tsallis statistics \cite{Tsallis:1988:Generalization,Tsallis:2009:NonextensiveBook}.
It was shown that long-range interactions or long-time memory effects
are responsible for such distributions. Since the noise intensity
of our model is governed by the Ornstein--Uhlenbeck process, correlation
time of the noise intensity serves as the memory effect, which is
not the case for the white noise.

\subsection{Relation to colored noise\label{sub:relation_to_colored_noise}}

As argued in Sec.~\ref{sec:introduction}, colored noise is a generalization
of white noise and also incorporates mesoscopic time-scale inhomogeneity.
However, many approximation schemes for colored noise, including adiabatic
elimination, can take into account a time-correlation effect {[}\emph{i.e.},
$O(\tau)${]} within the conventional FPE. The necessity of derivatives
of higher order than two in our systems indicates that systems driven
by SIN are essentially different than those governed by colored noise.

For colored noise driven systems, one popular approach to obtain approximate
FPE is the unified colored-noise approximation (UCNA), which is based
on the adiabatic elimination approach \cite{Jung:1987:ColoreApproximation}.
In Refs.~\cite{Colet:1989:ColorNoisePI,Wio:1989:PItoColoredNoise},
UCNA was confirmed as a reliable Markovian approximation by the path-integral.
In Refs.~\cite{Colet:1989:ColorNoisePI,Wio:1989:PItoColoredNoise},
they derived FPE using a Markovian Lagrangian function in the path-integral
representation, where non-Markovian terms, such as $\ddot{x}$ and
$\dot{x}^{n}$ ($n\ge2$), were ignored. The path-integral was applied
to non-Gaussian colored noise in Ref.~\cite{Fuentes:2002:EFPE},
and an effective Markovian FPE was obtained in this case. Since our
obtained equation includes derivatives of higher order than two, a
naive application of the path-integral techniques used in the colored
noise case may not be possible. However, it is important to understand
our approximation scheme from the viewpoint of the path-integral,
because the path-integral has an intuitive interpretation in stochastic
processes. We leave these for future studies.

\subsection{Cases of the other $\gamma$ range\label{sub:other_range}}

\begin{figure}
\begin{centering}
\includegraphics[width=6cm]{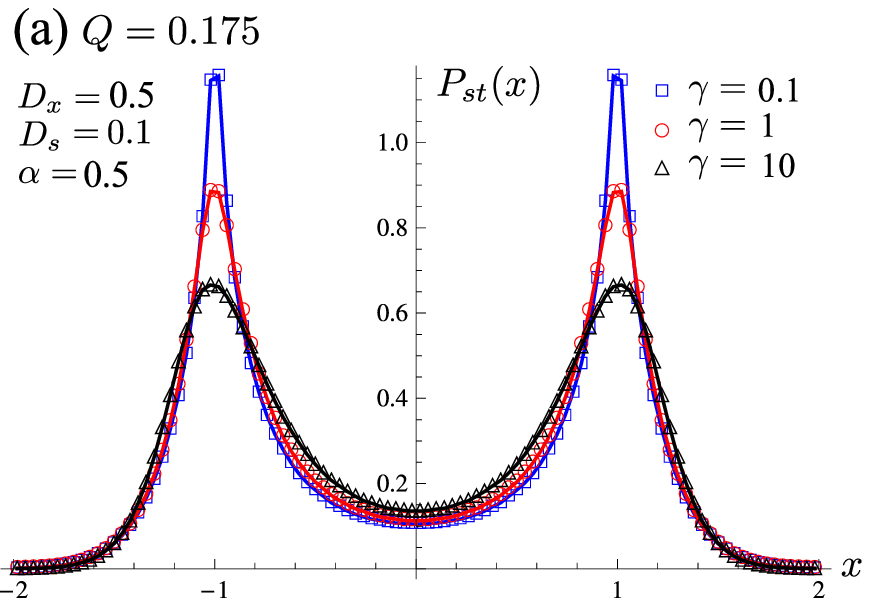}~~~~\includegraphics[width=6cm]{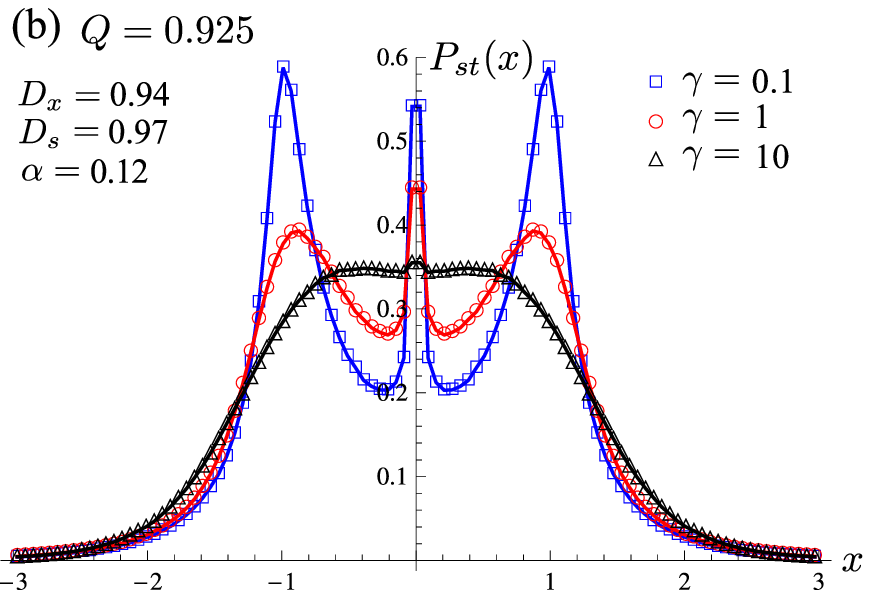} 
\par\end{centering}

\caption{(Color online) Stationary distributions of the quartic bistable potential
with (a) additive and (b) multiplicative noise for $\gamma=0.1$ (squares),
$1$ (circles) and $10$ (triangles). Parameter values are (a) $D_{x}=0.5$,
$D_{s}=0.1$, and $\alpha=0.5$ ($Q=0.175$) and (b) $D_{x}=0.94$,
$D_{s}=0.97$, and $\alpha=0.12$ ($Q=0.925$). These data are from
MC simulations with $N=5\times10^{7}$ samples. Solid lines are included
as a guide to the eye only. \label{fig:other_MC}}

\end{figure}

In preceding sections, we described the approximation scheme for the
$\gamma\gg1$ case. It is important to consider the other $\gamma$
range. In this section, we describe MC simulations carried out for
$\gamma=0.1$, $1$, and $10$ for two settings: the quartic bistable
potential with additive and with multiplicative noise. Parameters
($D_{x}$, $D_{s}$, and $\alpha$) of the additive and multiplicative
noise cases are identical to those shown in Figs.~\ref{fig:quartic_comp}(b)
and \ref{fig:quartic_mult_comp}(c), respectively. These results are
shown in Fig.~\ref{fig:other_MC}. Fig.~\ref{fig:other_MC}(a) shows
that higher peaks emerge at $x=\pm1$ for smaller $\gamma$. For Fig.~\ref{fig:other_MC}(b),
densities at $x=-1$, $0$, and $1$ are higher for smaller $\gamma$.
Similar trends have also been observed in Secs.~\ref{sub:p_q_additive}
and \ref{sub:quartic_mult}. Results of Sec.~\ref{sub:quartic_mult}
and this section show that trimodality is induced when the noise-intensity
fluctuations are meso/macroscopic and the fluctuation width is large.
The stationary distributions of smaller $\gamma$ can be explained
by the superposition of different distributions. For $\gamma\rightarrow0$,
the case reduces to that of superstatistics, where the stationary
distribution is given in a Bayesian fashion: \[
P_{st}(x)\sim\int ds\, P_{st}(x|s)P(s).\]
 Here, $P_{st}(x|s)$ is the stationary distribution taking $s$ as
a parameter, and $P(s)$ is the distribution of $s$.

Ref.~\cite{Queiros:2008:SSMultiplicative} considered a different
superstatistical model of multiplicative noise processes:\[
\frac{dx}{dt}=f(x)+[x^{2}]^{\alpha}\xi(t),\]
 which assumes that multiplicative noise exponent $\alpha$ fluctuates
{[}$\xi(t)$ is white noise{]}. Stationary distributions of their
model also exhibit the transitions of stationary distributions.

\section{Concluding Remarks\label{sec:conclusion}}

\begin{table}
\caption{RMS distance between HFPE (or FPE) and MC distributions for parabolic
and quartic bistable potentials with additive noise, quartic potential
with multiplicative noise, and a gene expression model. \label{tab:RMS}}

\medskip{}

\begin{centering}
\begin{tabular}{|c|c|c|c|c|c|c|}
\hline 
 & \multicolumn{3}{c|}{parabolic potential} & \multicolumn{3}{c|}{quartic bistable potential}\tabularnewline
 & \multicolumn{3}{c|}{with additive noise} & \multicolumn{3}{c|}{with additive noise}\tabularnewline
\cline{2-7} 
 & Fig.~2(a) & Fig.~2(b) & Fig.~2(c) & Fig.~3(a) & Fig.~3(b) & Fig.~3(c)\tabularnewline
\hline
\hline 
HPFE & 0.00418 & 0.00788 & 0.0934 & 0.0120 & 0.0173 & 0.0253\tabularnewline
\hline 
FPE & 0.0184 & 0.0229 & 0.152 & 0.0338 & 0.0267 & 0.0329\tabularnewline
\hline
\end{tabular}
\par\end{centering}

\medskip{}

\centering{}\begin{tabular}{|c|c|c|c|c|}
\hline 
 & \multicolumn{3}{c|}{quartic bistable potential} & gene expression\tabularnewline
 & \multicolumn{3}{c|}{with multiplicative noise} & model\tabularnewline
\cline{2-5} 
 & Fig.~4(a) & Fig.~4(b) & Fig.~4(c) & Fig.~6\tabularnewline
\hline
\hline 
HPFE & 0.00683 & 0.00647 & 0.0143 & 0.00446\tabularnewline
\hline 
FPE & 0.0165 & 0.0113 & 0.0181 & 0.0163\tabularnewline
\hline
\end{tabular}
\end{table}

In this paper, we have derived the time evolution equation of systems
driven by SIN using adiabatic elimination under the assumption $\gamma\gg1$.
The obtained HFPE is of $O(\gamma^{-1})$ and contains derivatives
up to fourth order. We have calculated the stationary distributions
of the obtained equation by using the perturbation expansion and applied
them to three different cases. Table \ref{tab:RMS} summarizes the
RMS distance between the stationary distributions calculated by the
HFPE (or FPE) and MC which have been shown in Figs. \ref{fig:quadratic_comp},
\ref{fig:quartic_comp}, \ref{fig:quartic_mult_comp} and \ref{fig:gene_expression}.
It clearly shows that the results of the HFPE are better than those
of FPE. We have pointed out that SIN makes densities at stable sites
higher. Furthermore, SIN induces the transition from unimodal to bimodal
in the quartic bistable potential driven by multiplicative noise.
For specific parameters, the stationary distribution exhibits three
peaks, which are not observed in conventional white noise cases. Because
the stationary distributions obtained by Eqs.~(\ref{eq:WKB_DE1})
and~(\ref{eq:WKB_DE2}) are given by the first-order differential
equations, their analytical expressions can be calculated for many
general configurations. The applicability has been shown with the
stochastic gene expression model, which has a non-trivial drift term.
This indicates that our method can be applied to many real-world phenomena.
Furthermore, higher-order derivatives in HFPE have been given much
attention in recent years \cite{Plyukhin:2005:GFPE,Plyukhin:2008:GFPE}.
Analysis of higher-order terms in our systems are also important,
and so we intend to pursue this in future studies.

\section*{Acknowledgments}

This work was supported by a Grant-in-Aid for Scientific Research
on Priority Areas {}``Systems Genomics'' from the Ministry of Education,
Culture, Sports, Science and Technology, Japan.

\appendix
%dummy comment inserted by tex2lyx to ensure that this paragraph is not empty
%dummy comment inserted by tex2lyx to ensure that this paragraph is not empty

\section{Derivation of the Higher-Order Fokker--Planck Equation\label{sec:derivation_GFPE}}

In this appendix, we explain the procedures of adiabatic elimination
with eigenfunction expansion used in our model, following Refs.~\cite{Kaneko:1976:AdiabaticElim,Risken:1989:FPEBook}.

We first expand $P(x,s;t)$ into the complete set $\psi_{n}(s)$ of
the operator $\mathrm{L}_{s}(s)$ {[}Eq.~(\ref{eq:Ls}){]}: \begin{equation}
P(x,s;t)=\sum_{n=0}^{\infty}P_{n}(x;t)\psi_{n}(s),\label{eq:ef_expansion}\end{equation}
 where $\psi_{n}(s)$ are eigenfunctions of the following equation:
\[
\mathrm{L}_{s}(s)\psi_{n}(s)=-\lambda_{n}\psi_{n}(s).\]
 Eigenvalues $\lambda_{n}$ and eigenfunctions $\psi_{n}(s)$ are
given by \begin{equation}
\lambda_{n}=n,\label{eq:eigenvalue}\end{equation}

\begin{equation}
\psi_{n}(s)=\sqrt{\frac{1}{2\pi D_{s}}}\frac{1}{2^{n}n!}H_{n}\left(\eta\right)\exp\left(-\eta^{2}\right),\label{eq:def_eigenfunction}\end{equation}
 where $\eta=\sqrt{1/(2D_{s})}(s-\alpha)$, and $H_{n}(\eta)$ is
the $n$th Hermite polynomial. We introduce the adjoint function $\psi_{n}^{\dagger}(s)$
of $\psi_{n}(s)$: \begin{equation}
\psi_{n}^{\dagger}(s)=H_{n}\left(\eta\right),\hspace{1em}\psi_{0}^{\dagger}(s)=1.\label{eq:def_adjoint}\end{equation}
 The orthonormality and complete relations are\begin{equation}
\left\langle \psi_{n}^{\dagger}(s)\psi_{m}(s)\right\rangle =\delta_{nm},\label{eq:orthonormality}\end{equation}
 where $\left\langle \cdots\right\rangle $ denotes an integration
with respect to $s$ {[}$\left\langle A(s)\right\rangle =\int ds\, A(s)$
{]}. Using Eqs.~(\ref{eq:def_adjoint}) and~(\ref{eq:orthonormality}),
$P(x,t)$ ($=\left\langle P(x,s,t)\right\rangle $ ) is given by \begin{equation}
P(x;t)=\sum_{n=0}^{\infty}P_{n}(x;t)\left\langle \psi_{0}^{\dagger}(s)\psi_{n}(s)\right\rangle =P_{0}(x;t).\label{eq:PDF_xt}\end{equation}
 By multiplying both sides of Eq.~(\ref{eq:FPE}) by $\psi_{m}^{\dagger}(s)$
and integrating out $s$, we obtain {[}$P_{m}=P_{m}(x;t)${]}\begin{equation}
\frac{\partial}{\partial t}P_{m}=-\frac{\partial}{\partial x}fP_{m}+D_{x}\sum_{n=0}^{\infty}\left\langle \psi_{m}^{\dagger}s^{2}\psi_{n}\right\rangle \triangle_{g}P_{n}-\gamma\lambda_{m}P_{m},\label{eq:recursive_Eq}\end{equation}
 where $\triangle_{g}$ is an operator defined by Eq.~(\ref{eq:def_op}).

Using relations of the Hermite polynomial, specifically, $H_{n+1}(z)=2zH_{n}(z)-2nH_{n-1}(z)$
and $H_{0}(z)=1$, the following relation holds: \begin{eqnarray}
\left\langle \psi_{m}^{\dagger}(s)s^{2}\psi_{n}(s)\right\rangle  & = & 2D_{s}m(m-1)\delta_{m-2,n}+2\sqrt{2D_{s}}\alpha m\delta_{m-1,n}+\frac{1}{2}D_{s}\delta_{m+2,n}\nonumber \\
 &  & +\sqrt{2D_{s}}\alpha\delta_{m+1,n}+\left\{ 2D_{s}\left(m+\frac{1}{2}\right)+\alpha^{2}\right\} \delta_{m,n}.\label{eq:Hermite_relation}\end{eqnarray}
 According to Eqs.~(\ref{eq:recursive_Eq}) and~(\ref{eq:Hermite_relation}),
time evolution of $P_{0}$ is governed by\begin{equation}
\frac{\partial}{\partial t}P_{0}=-\frac{\partial}{\partial x}fP_{0}+D_{x}(D_{s}+\alpha^{2})\triangle_{g}P_{0}+\sqrt{2D_{s}}\alpha D_{x}\triangle_{g}P_{1}+\frac{1}{2}D_{s}D_{x}\triangle_{g}P_{2}.\label{eq:P0_relation}\end{equation}
 Equation~(\ref{eq:P0_relation}) includes $P_{1}$ and $P_{2}$,
although we want to obtain a closed equation of $P_{0}$. We calculate
$P_{m}$ ($m\ge1$) by Eq.~(\ref{eq:recursive_Eq}): \begin{equation}
P_{m}=\frac{1}{\lambda_{m}\gamma}\left\{ -\frac{\partial}{\partial x}fP_{m}+D_{x}\sum_{n=0}^{\infty}\left\langle \psi_{m}^{\dagger}s^{2}\psi_{n}\right\rangle \triangle_{g}P_{n}\right\} \hspace{1em}(\mathrm{for}\; m\ge1).\label{eq:Pm}\end{equation}
 In Eq.~(\ref{eq:Pm}), we ignore a time derivative term. By substituting
Eq.~(\ref{eq:Pm}) into Eq.~(\ref{eq:P0_relation}), up to $O(\gamma^{-1})$,
we obtain Eq.~(\ref{eq:KM_expansion_ori}).

\end{document}